# 4Cast Lab: Open-Source App for Machine Learning & Statistical Analysis

**Periklis Gogas**
Department of Economics
Democritus University of Thrace
Komotini 69100, Greece
pgkogkas@econ.duth.gr

## Abstract

4Cast Lab is a free, open-source application for machine learning, econometrics and statistical analysis that runs on the user's own computer, with no installation, no programming and no data leaving the machine. It was developed at the E|M|E|F Lab, Department of Economics, Democritus University of Thrace, for researchers, analysts, instructors and students in any field. Its Forecasting mode trains eight learning algorithms (with three kernels for SVM/SVR), implemented from scratch, under leakage-safe validation for time-ordered data; it selects models by cross-validation, compares the selected model with every alternative and with naive benchmarks through formal tests of predictive accuracy, and documents each run in a methods report and a Python script that the user can export. Its Estimation mode covers the core of applied econometrics, from OLS with robust inference to ARIMA, ARDL, VAR/VECM and GARCH. This paper describes the design, the interface and the workflow of version 4.6.2, reports a validation of 1,234 numerical results against statsmodels, linearmodels, arch, scikit-learn and NIST certified values, illustrates the workflow with three examples, and discusses current limitations.

**Download:** https://github.com/PerryGogas/4Cast-Lab/releases/latest

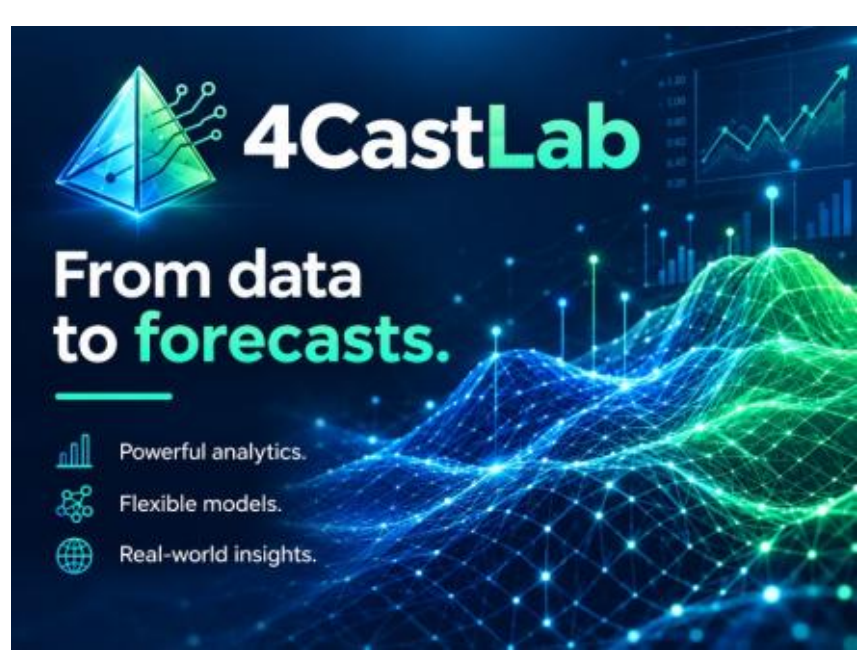

# 1. Introduction

Machine learning and statistical modelling are now standard tools in teaching, research and practice across many fields, yet using them well meets two obstacles. The first is technical: users without programming experience (especially students) spend much of their time installing software and debugging code rather than on modelling choices. The second is methodological: general-purpose tools rarely warn against the errors that matter most in real data, such as look-ahead bias, model selection on the test sample, and claims of forecasting success that actually do not beat a random walk.

4Cast Lab addresses both. It opens as a single file and guides the user from data import to a documented, tested model. Its defaults follow the forecasting literature rather than generic machine learning practice: time-ordered splits, selection by cross-validation, benchmarks against naive forecasts, and significance tests of predictive accuracy. The aim is that a user who follows the defaults produces results that would survive a referee's first questions.

The paper makes three contributions. It describes the design and interface of the software (Sections 2 and 3). It reports a systematic validation of the numerical results against established reference software (Section 4) which is a precondition for using any program in theses and published research. Finally, it illustrates the workflow with three examples that show what the software reports and how to read it (Section 5).

# 2. Design and architecture

4Cast Lab is a single HTML file written in JavaScript. All learning algorithms and estimators are implemented from scratch, so no server, Python environment or package manager is needed. Chart.js draws the figures and SheetJS reads and writes Excel files. The same file is compiled into a portable Windows executable that runs without installation or administrator rights. Four design choices shape the app:

- **Privacy by construction.** All computation is local. Nothing is uploaded, which allows the use of confidential, proprietary or licensed data. The only network connections are an update check and, if the user chooses to use it, an optional AI interpretation of the results through a provider of the user's choice.
- **Two views.** A Core view shows the essential settings, each with a short explanation; an Expert view exposes the full set of options and abilities of the app, and the Estimation mode. Switching views only shows or hides panels: the results are identical in both.
- **Guidance in the interface.** Every panel carries a "How to read this" note, every setting a help icon, and every set of results an "In plain words" summary that states the conclusion in one paragraph, including when a model fails to beat its benchmark. A Glossary tab defines every metric and test.

- **Reproducibility.** Every run records its settings and random seed, can be saved as a session file (.4cl) and reopened, and can be exported as a pandas and scikit-learn script that replays the split, preprocessing, model and metrics. A Methods Report writes the prose for the Methods section of a paper or thesis, and a Full Report documents the analysis from start to end.

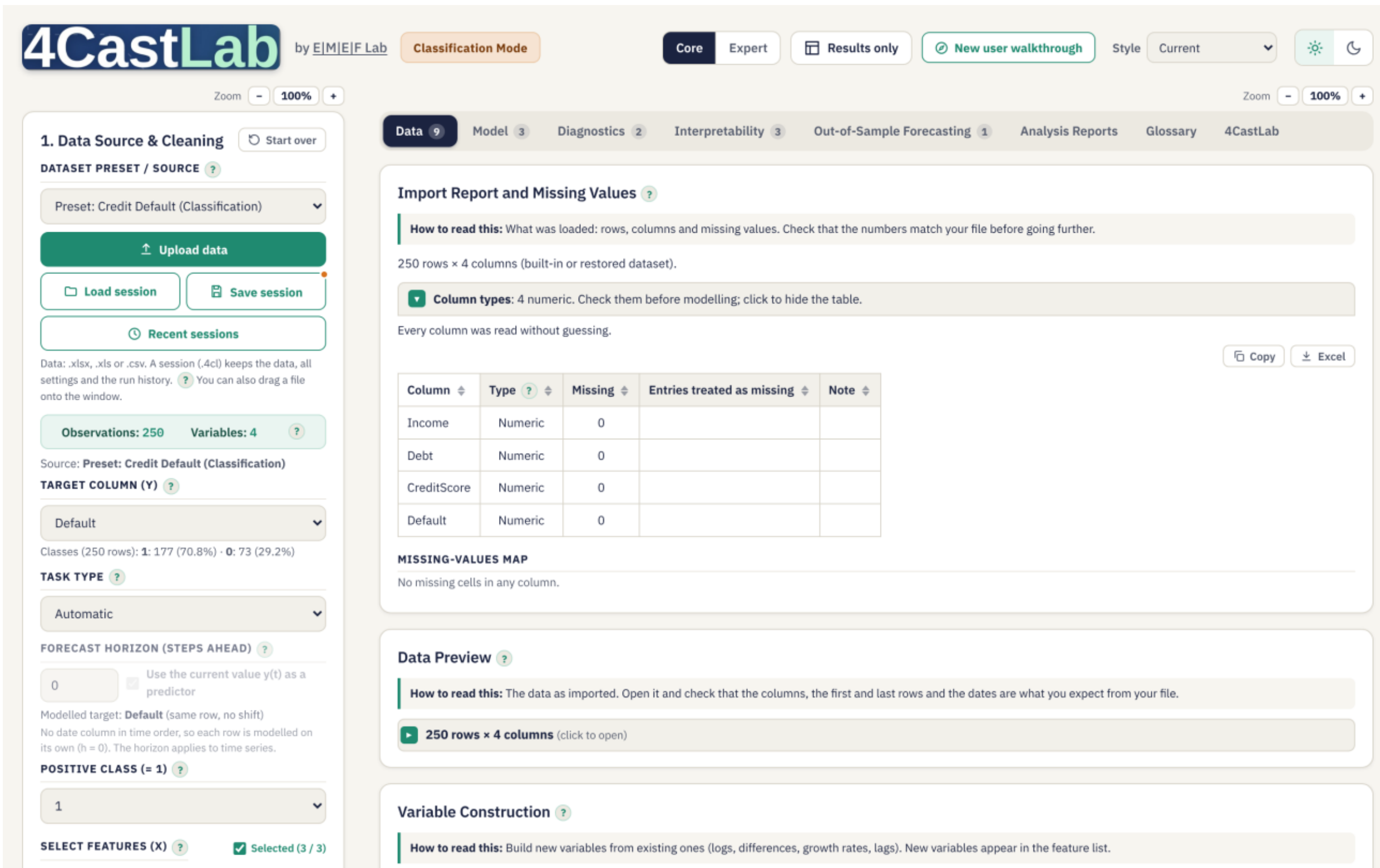

**Figure 1.** The Core view (Forecasting mode). Settings are on the left; results appear in the tabs on the right, each with a "How to read this" note.

# 3. Methods and workflow

## 3.1 Forecasting (machine learning) mode

A run in Forecasting mode follows six steps, each on its own tab:

1. **Data.** Import (or drag and drop) CSV or Excel files, or use the built-in examples. An import report and a data preview list the column types, missing values, date coverage, gaps and duplicates. The app flags possible target leakage, checks multicollinearity (VIF) and supports panel data through a unit column.
2. **Target and horizon.** For forecasting, the target is shifted to y(t+h), with h = 1 by default. Lags, differences and transformations are computed within each unit.
3. **Validation design.** A time-series split with an optional embargo, and expanding-window cross-validation that purges rows around each fold (López de Prado, 2018), to avoid data leakage: no future information enters the training set. Random, stratified splits are available for cross-sections, with a warning when they are used on dated data.

4. **Training and tuning.** Eight algorithms are available, individually or all at once (*Train All Models & Compare*); Table 1 lists them. Hyperparameters are tuned by cross-validation on the training set only, with a quick grid, full research grids, exhaustive grid search or random search (Bergstra and Bengio, 2012), and optional nested cross-validation. Variable selection offers LASSO, elastic net, recursive feature elimination, stability selection and filter methods.
5. **Evaluation.** Regression reports RMSE, MAE, MAPE, sMAPE, $R^2$ and MASE (Hyndman and Koehler, 2006), with conformal prediction intervals (Vovk, Gammerman and Shafer, 2005). Classification reports ROC-AUC, PR-AUC, accuracy, sensitivity, specificity, F1, MCC, the Brier score and log loss, and adds class-imbalance handling, probability calibration (Platt scaling or isotonic regression) and cost-based thresholds. Every metric is shown for a) the cross-validation folds, b) the whole training rows and c) the Test Set, and are kept apart.
6. **Interpretation and reporting.** Permutation importance (Fisher, Rudin and Dominici, 2019), Shapley values (Štrumbelj and Kononenko, 2010), impurity importance, mutual information and correlation rank the predictors; their agreement is measured by Kendall's W (Kendall and Babington Smith, 1939). Partial-dependence and ICE curves show the shape of each effect. Forecasts for new data and multi-step forecasts beyond the sample (direct or iterated) come with conformal or simulated bands.

**Table 1.** Learners in Forecasting mode

| Family | Learners | Notes |
|---|---|---|
| Linear | Linear regression (OLS); logistic regression | exact QR least squares |
| Instance-based | k-nearest neighbours | k tuned by CV |
| Trees and ensembles | Decision tree, random forest, gradient boosting, stacking | Breiman (2001); Friedman (2001) |
| Kernel | Support vector machines (classification), SVR and kernel ridge regression | linear, RBF and polynomial kernels |
| Neural | Feed-forward network (MLP) | ReLU, sigmoid or tanh; early stopping |

Eight algorithms in five families; with three kernels, *Train All* compares ten configurations.

**Kernels.** The support vector machine (Cortes and Vapnik, 1995) algorithm offers linear, RBF (Gaussian) and polynomial kernels, with the kernel treated as a hyperparameter in *Train All*. Classification is solved exactly by sequential minimal optimisation (Platt, 1998), as in LIBSVM (Chang and Lin, 2011) and scikit-learn, with a faster approximate solver as an alternative; the kernel width defaults to the scikit-learn "scale" rule, $\gamma = 1/(p \cdot \mathrm{Var}\ X)$. For regression the linear kernel uses support vector regression, and the RBF and polynomial kernels use kernel ridge regression in closed form. Because non-linear kernels need an $n \times n$ Gram matrix, the app warns when the training sample is large enough to make them slow.

### 3.2 Benchmarks and the evaluation of the selected model

The central device is the benchmark. On time-ordered data every model is compared with the stronger of the random walk and the training mean, chosen on training rows only. The Diebold and Mariano (1995) test with the Harvey, Leybourne and Newbold (1997) correction judges whether the difference in accuracy is significant; the Clark and West (2007) test allows for the benchmark being nested in the model; and the Pesaran and Timmermann (1992) test checks directional accuracy. For classification the benchmark is the base-rate forecast, and the AUC is tested against 0.5 (DeLong, DeLong and Clarke-Pearson, 1988).

Model selection is kept separate from evaluation. *Train All* gives each algorithm its own cross-validated grid search and selects the model with the best mean cross-validation score; the Test Set is used only once, to evaluate the selected model on data that no model has seen. A second table then asks whether the selected model is really better: it compares the selected model with each other model and with the benchmark on the same Test Set rows, using the Diebold-Mariano, DeLong or McNemar (1947) test according to the selection metric, a paired bootstrap interval of the difference, and Holm's (1979) correction for the number of comparisons. These tests check the choice made by cross-validation; they never change it, so the Test Set does not leak into the selection (Figure 2).

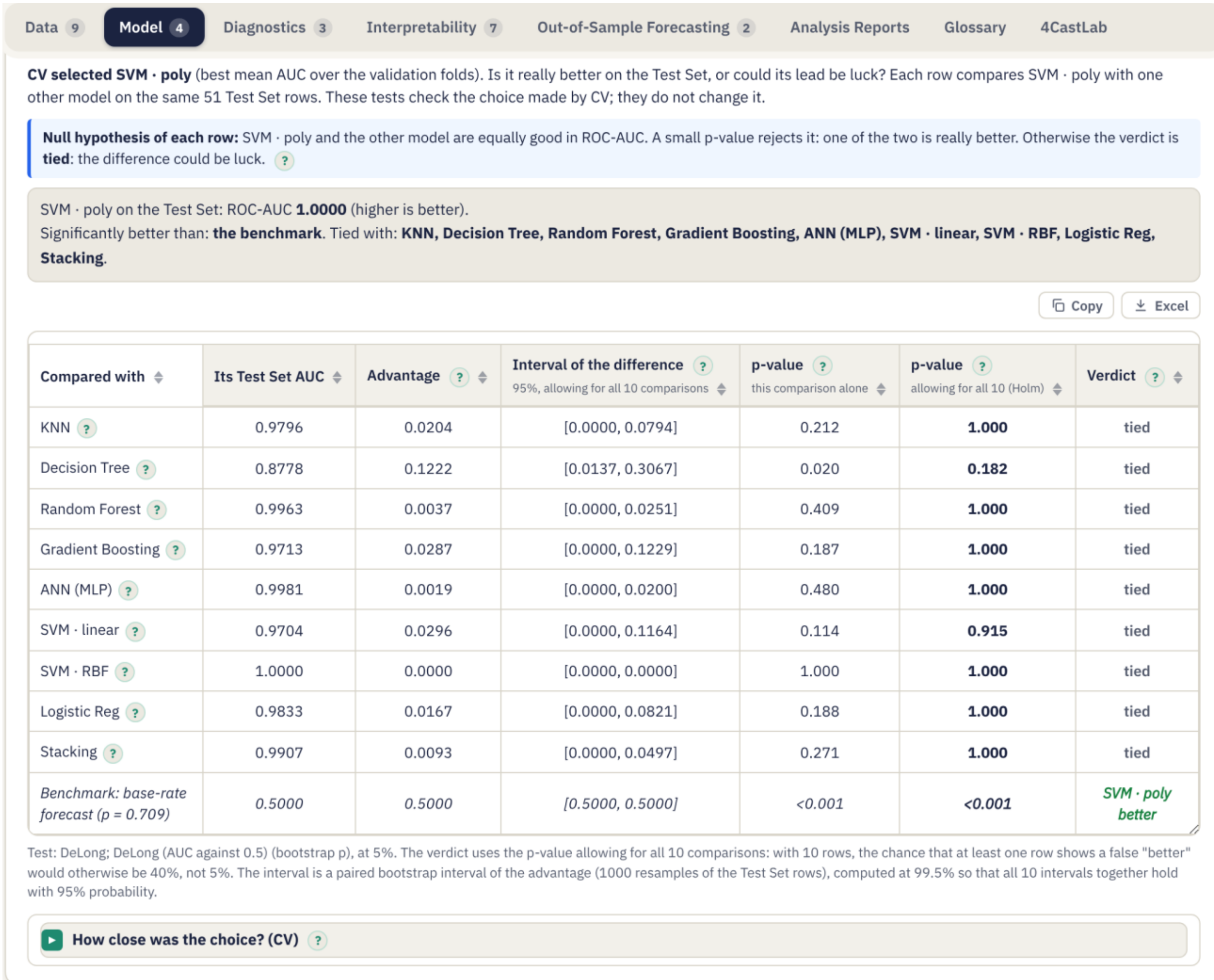

Data 9 | Model 4 | Diagnostics 3 | Interpretability 7 | Out-of-Sample Forecasting 2 | Analysis Reports | Glossary | 4CastLab

**CV selected SVM · poly** (best mean AUC over the validation folds). Is it really better on the Test Set, or could its lead be luck? Each row compares SVM · poly with one other model on the same 51 Test Set rows. These tests check the choice made by CV; they do not change it.

**Null hypothesis of each row:** SVM · poly and the other model are equally good in ROC-AUC. A small p-value rejects it: one of the two is really better. Otherwise the verdict is **tied**: the difference could be luck. ?

SVM · poly on the Test Set: ROC-AUC **1.0000** (higher is better).
Significantly better than: **the benchmark**. Tied with: **KNN, Decision Tree, Random Forest, Gradient Boosting, ANN (MLP), SVM · linear, SVM · RBF, Logistic Reg, Stacking**.

Copy | Excel

| Compared with | Its Test Set AUC | Advantage ? | Interval of the difference ? 95%, allowing for all 10 comparisons | p-value ? this comparison alone | p-value ? allowing for all 10 (Holm) | Verdict ? |
|---|---|---|---|---|---|---|
| KNN ? | 0.9796 | 0.0204 | [0.0000, 0.0794] | 0.212 | **1.000** | tied |
| Decision Tree ? | 0.8778 | 0.1222 | [0.0137, 0.3067] | 0.020 | **0.182** | tied |
| Random Forest ? | 0.9963 | 0.0037 | [0.0000, 0.0251] | 0.409 | **1.000** | tied |
| Gradient Boosting ? | 0.9713 | 0.0287 | [0.0000, 0.1229] | 0.187 | **1.000** | tied |
| ANN (MLP) ? | 0.9981 | 0.0019 | [0.0000, 0.0200] | 0.480 | **1.000** | tied |
| SVM · linear ? | 0.9704 | 0.0296 | [0.0000, 0.1164] | 0.114 | **0.915** | tied |
| SVM · RBF ? | 1.0000 | 0.0000 | [0.0000, 0.0000] | 1.000 | **1.000** | tied |
| Logistic Reg ? | 0.9833 | 0.0167 | [0.0000, 0.0821] | 0.188 | **1.000** | tied |
| Stacking ? | 0.9907 | 0.0093 | [0.0000, 0.0497] | 0.271 | **1.000** | tied |
| *Benchmark: base-rate forecast (p = 0.709)* | *0.5000* | *0.5000* | *[0.5000, 0.5000]* | *<0.001* | ***<0.001*** | *SVM · poly better* |

Test: DeLong; DeLong (AUC against 0.5) (bootstrap p), at 5%. The verdict uses the p-value allowing for all 10 comparisons: with 10 rows, the chance that at least one row shows a false "better" would otherwise be 40%, not 5%. The interval is a paired bootstrap interval of the advantage (1000 resamples of the Test Set rows), computed at 99.5% so that all 10 intervals together hold with 95% probability.

▶ How close was the choice? (CV) ?

**Figure 2.** "Is the model selected by CV really better?": pairwise tests of the selected model against every other model and the benchmark on the Test Set, with Holm-adjusted p-values and bootstrap intervals.

## 3.3 Estimation (econometrics) mode

Estimation mode, in the Expert view, fits models to the whole sample for explanation and inference, as in Stata or EViews (Figure 3). It offers OLS with classical, White (1980) HC0–HC3, Newey and West (1987) HAC and clustered standard errors; logit and probit with average marginal effects; IV/2SLS with first-stage diagnostics; ARIMA and SARIMA by exact maximum likelihood; ARDL with the bounds test of Pesaran, Shin and Smith (2001); error-correction models; VAR with impulse responses and variance decompositions; Johansen (1991) cointegration and VECM; and GARCH (Bollerslev, 1986). A specification box adds dummies, interactions, polynomials, lags, trends and seasonal dummies. The Tests tab reports residual diagnostics, specification tests, Wald tests of any linear restriction, multicollinearity and structural breaks (Chow, Bai-Perron), each with a plain reading at the 5% level. Estimated models can be collected in a journal-style Results Table exported to LaTeX, Word or Excel (Figure 4), and an ARIMA, ARDL or VAR can be sent to Forecasting mode as a benchmark for the machine learning models on the same Test Set rows.

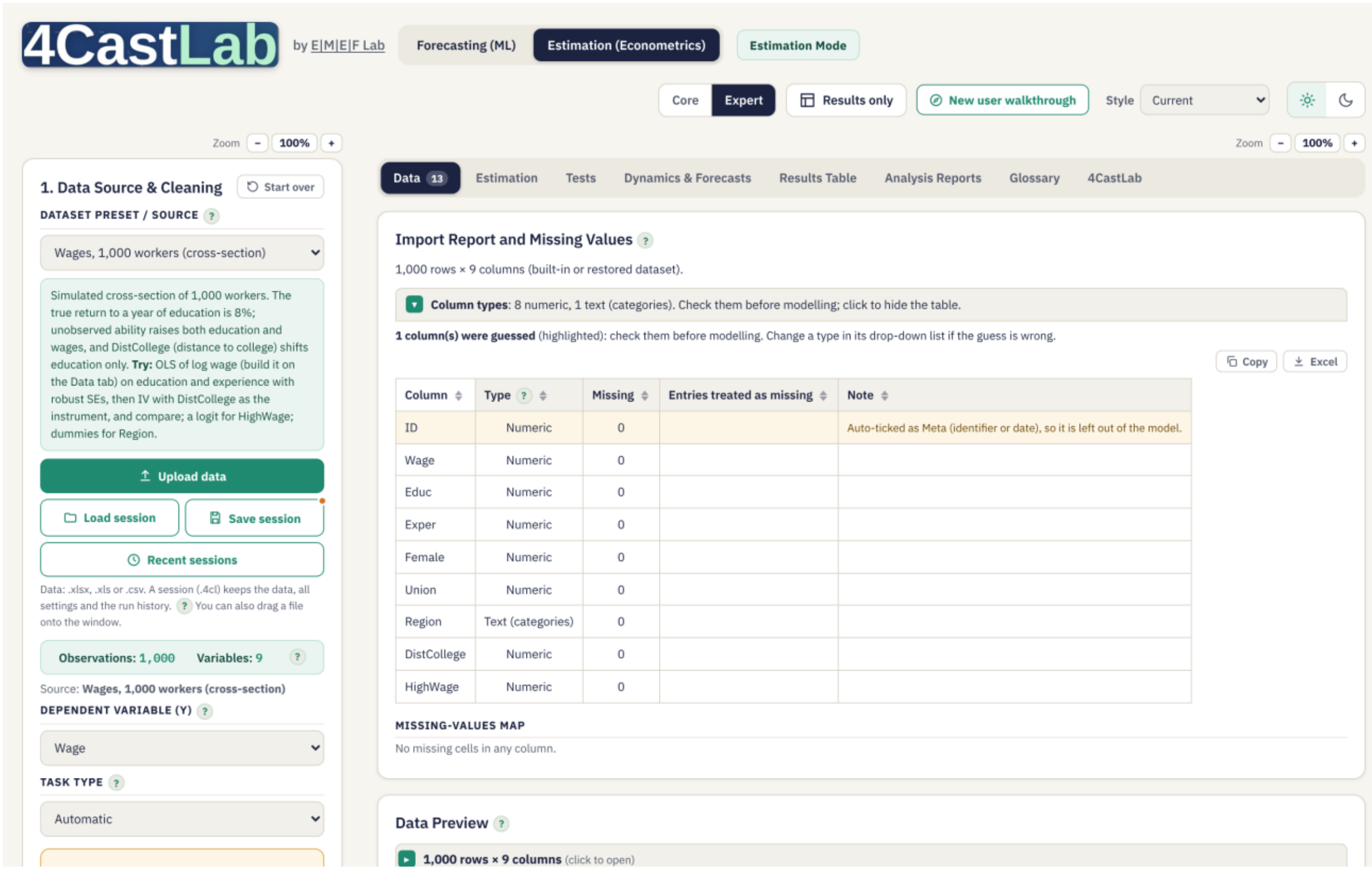

**Figure 3.** The Expert view with the Estimation (Econometrics) mode selected and the Wages example loaded.

# 4. Validation

Software used for theses and published research must reproduce established results. Version 4.6.2 was therefore tested against reference implementations on standard public datasets and certified benchmarks. Each test calls the app's own numerical routines, or drives its interface in a headless browser, and compares the output with the reference on identical inputs: statsmodels 0.15 (Seabold and Perktold, 2010), linearmodels 7.0, arch 8.0, scikit-learn 1.9 (Pedregosa et al., 2011), the NIST StRD certified values for the Longley (1967) data, and R where statsmodels follows a different convention. Table 2 summarises the 1,234 checks.

**Table 2.** Validation of 4Cast Lab 4.6.2 against reference software

| Area | Reference | Datasets | Checks |
|---|---|---|---|
| Linear models (OLS, robust/HAC/cluster SEs, tests) | NIST StRD, statsmodels | Longley; Wooldridge wage1; Grunfeld; US macrodata | 78 |
| Logit and probit, marginal effects | statsmodels | Spector–Mazzeo; Wooldridge mroz | 65 |
| IV / 2SLS | linearmodels | Wooldridge card and mroz | 30 |
| Unit-root and series tests | statsmodels, arch | US macrodata series | 386 |
| ARIMA / SARIMA | statsmodels, R | AirPassengers; US macrodata | 48 |
| ARDL, bounds test, ECM | statsmodels | Danish money demand | 81 |
| VAR, Johansen, VECM | statsmodels | US macrodata | 179 |
| GARCH | arch | S&P 500 daily returns | 30 |

| Area | Reference | Datasets | Checks |
|---|---|---|---|
| Forecast-comparison tests | arch, published formulas | Breast cancer; US GDP growth | 88 |
| Machine learning models | scikit-learn | Breast cancer; diabetes; Friedman #1; two moons | 159 |
| Splits, cross-validation, interpretability | independent code | synthetic designs | 56 |
| Interface, exports and sessions | values above | end-to-end runs | 34 |
| **Total** | | | **1,234** |

All 1,234 checks pass. Deterministic results agree within a relative difference of $10^{-6}$ (1,168 checks; 1,145 within $10^{-9}$); iterative and stochastic estimators are compared within stated tolerances.

The application passes all checks. The deterministic estimators (least squares, maximum likelihood with closed-form or well-conditioned optima, tests and their p-values) agree with the references to within $10^{-6}$ in relative terms, most to $10^{-9}$ or better, including the ill-conditioned Longley regression. Random forest, gradient boosting and the neural network cannot match scikit-learn exactly, because their random number streams and solvers differ; they were checked for equal accuracy on the same rows instead and passed.

Where the app and a reference disagree because they follow different conventions, the app follows the convention used in the econometric literature and in Stata, and the Methods Report states it. Examples are the regression form of the Wu-Hausman test, $t(G-1)$ inference with clustered standard errors, the HC1 small-sample factor, and the textbook form of the Johansen test with a trend. In four cases the reference itself was at fault: an off-by-one count of regressors in the critical values of the statsmodels bounds test, an approximation in the statsmodels Johansen test with a trend, a projection variant of the Wu-Hausman test in linearmodels, and printed values for the Box-Jenkins airline model that come from an older estimator.

The 1,234 checks also revealed six defects in a pre-release build, all corrected in the released version 4.6.2: a crash in the GARCH starting values for 22 of 36 model orders; standard errors that became undefined when a GARCH parameter lay on its bound; a local optimum for GJR-GARCH(1,2,1); a small ridge penalty in the machine learning linear regression, now replaced by an exact QR solution; and two wording errors in the Methods Report about tuning and the SVM kernel. The complete validation report, with every check and its tolerance, and the test scripts are available from the author and with the release in Github.

# 5. Illustrative examples

The three examples below use the built-in teaching datasets, so any reader can reproduce them; all three are simulated with a known data-generating process. Each takes a few minutes and follows the Quick Start videos of the software.

## 5.1 Classification: credit default

The Credit Default example has 250 loan applicants, a binary target (default) and three predictors: income, debt and credit score. A stratified random split with a fixed seed keeps 51 applicants (20%) as the Test Set. A random forest with quick tuning reaches a cross-validated AUC of 0.978 and a Test Set AUC of 0.996, against 0.5 for the base-rate forecast (DeLong $p < 0.001$), and a Brier score of 0.047 against 0.208. *Train All* then tunes all ten configurations on the same rows (Table 3). Cross-validation selects the SVM with a polynomial kernel (C = 10; mean CV AUC 0.997, SE 0.002), whose Test Set AUC is 1.000. The evaluation of the selected model shows that it is significantly better than the benchmark but statistically tied with all nine alternatives after the Holm correction: on 51 rows, the data cannot separate the best models, and the app says so rather than simply declaring a winner. In the Interpretability tab, debt and credit score rank above income on most measures, with moderate agreement between the five criteria (Kendall's W = 0.52, p = 0.074).

**Table 3.** Credit default: Train All with quick tuning (cross-validation selects, the Test Set evaluates)

| Model | CV AUC | CV accuracy | Test AUC | Test accuracy |
|---|---|---|---|---|
| SVM, polynomial kernel (selected by CV) | 0.997 | 95.0% | 1.000 | 94.1% |
| SVM, RBF kernel | 0.994 | 95.0% | 1.000 | 98.0% |
| k-nearest neighbours | 0.988 | 93.0% | 0.980 | 94.1% |
| Stacking | 0.983 | 93.0% | 0.991 | 94.1% |
| Random forest | 0.978 | 91.0% | 0.996 | 94.1% |
| Neural network (MLP) | 0.970 | 90.5% | 0.998 | 98.0% |
| Gradient boosting | 0.964 | 89.9% | 0.971 | 94.1% |
| Logistic regression | 0.947 | 86.4% | 0.983 | 90.2% |
| SVM, linear kernel | 0.943 | 86.9% | 0.970 | 90.2% |
| Decision tree | 0.879 | 89.4% | 0.878 | 88.2% |

Training set 199 rows, Test Set 51 rows; 5-fold cross-validation on the training set. Models are ordered by CV AUC, the selection metric.

## 5.2 Forecasting: benchmark first

The Macro example has 288 simulated months (2000–2023) of industrial production, prices, money, the policy rate, stock returns and unemployment, which follows Okun's law. The target is next month's unemployment rate (h = 1) and the predictors are the other five series. The time-series split keeps the last 58 months as the Test Set, and each algorithm is tuned by expanding-window cross-validation on the first 229. Cross-validation selects gradient boosting (CV RMSE 0.358). Its Test Set $R^2$ is 0.87, which would look like a success in isolation. Table 4 shows why it is not: the random walk, chosen as the benchmark on the training rows, forecasts the Test Set better, with an RMSE of 0.183 against 0.224. On the training rows the ratio is 0.43, so the model's apparent advantage there is an artefact of fitting. The Diebold-Mariano test does not reject equal accuracy at 5% ($p = 0.054$), and the pairwise table finds gradient boosting tied with the random walk after the Holm correction ($p = 0.378$). The Clark-West test ($p = 0.032$) indicates that the predictors carry some information beyond the random walk, but not enough to produce more accurate forecasts on this sample, and directional accuracy is not significant (Pesaran-Timmermann $p = 0.380$). The "In plain words" summary states this conclusion directly.

**Table 4.** Unemployment forecasts, h = 1: the model selected by CV against the random walk

| | Gradient boosting | Random walk | Ratio |
|---|---|---|---|
| RMSE, training rows (in-sample) | 0.076 | 0.177 | 0.43 |
| RMSE, Test Set | 0.224 | 0.183 | 1.23 |
| MAE, Test Set | 0.178 | 0.146 | 1.22 |
| MASE, Test Set | 1.287 | 1.051 | 1.22 |
| $R^2$, Test Set | 0.873 | 0.916 | |
| Diebold-Mariano (HLN), squared loss | DM = 1.97, p = 0.054 | | |
| Clark-West (one-sided) | CW = 1.85, p = 0.032 | | |
| Directional accuracy (Pesaran-Timmermann) | 55.2%, p = 0.380 | | |

Training set 229 months, Test Set 58 months. Ratios below 1 favour the model. Selection by 5-fold expanding-window CV with quick tuning over ten configurations.

## 5.3 Econometrics: wage equation and ARIMA

The Wages example is a simulated cross-section of 1,000 workers. Table 5 reports the OLS regression of the hourly wage on education, experience, sex and union membership with White HC1 standard errors, as it appears in the Results Table. One more year of education raises the wage by 1.33 ($p < 0.001$), holding the other regressors constant. The Tests tab supports the choice of robust errors: the Koenker form of the Breusch-Pagan test rejects homoskedasticity (LM = 62.5, $p < 0.001$), while Breusch-Godfrey finds no serial correlation (p = 0.43). The RESET test (F = 10.7, $p < 0.001$) suggests that the level specification is too restrictive; in this simulated design wages are log-linear in education, which the user can verify by building log(wage) on the Data tab. The same dataset supports IV estimation with distance to college as the instrument.

**Table 5.** Wage equation, OLS with White HC1 standard errors

| | Coefficient | Std. error | p-value |
|---|---|---|---|
| Constant | −7.558 | 1.087 | < 0.001 |
| Education (years) | 1.329 | 0.083 | < 0.001 |
| Experience (years) | 0.090 | 0.011 | < 0.001 |
| Female | −1.403 | 0.292 | < 0.001 |
| Union member | 0.788 | 0.347 | 0.023 |
| $R^2$ / adjusted $R^2$ | 0.309 / 0.307 | | |
| Robust Wald F(4, 995) | 71.03 | | < 0.001 |
| Observations | 1,000 | | |

For time series, an ARIMA(1,1,1) of the monthly unemployment rate gives $\varphi = 0.732$ (SE 0.096) and $\theta = -0.429$ (SE 0.127), with an insignificant drift, by exact maximum likelihood (log-likelihood 114.46, AIC −220.9). The Dynamics & Forecasts tab reports forecasts for twelve months with 95% intervals that widen with the horizon, as they should for an integrated model (Figure 5). The same model can be sent to Forecasting mode and compared with the machine learning models of Section 5.2 on the same Test Set.

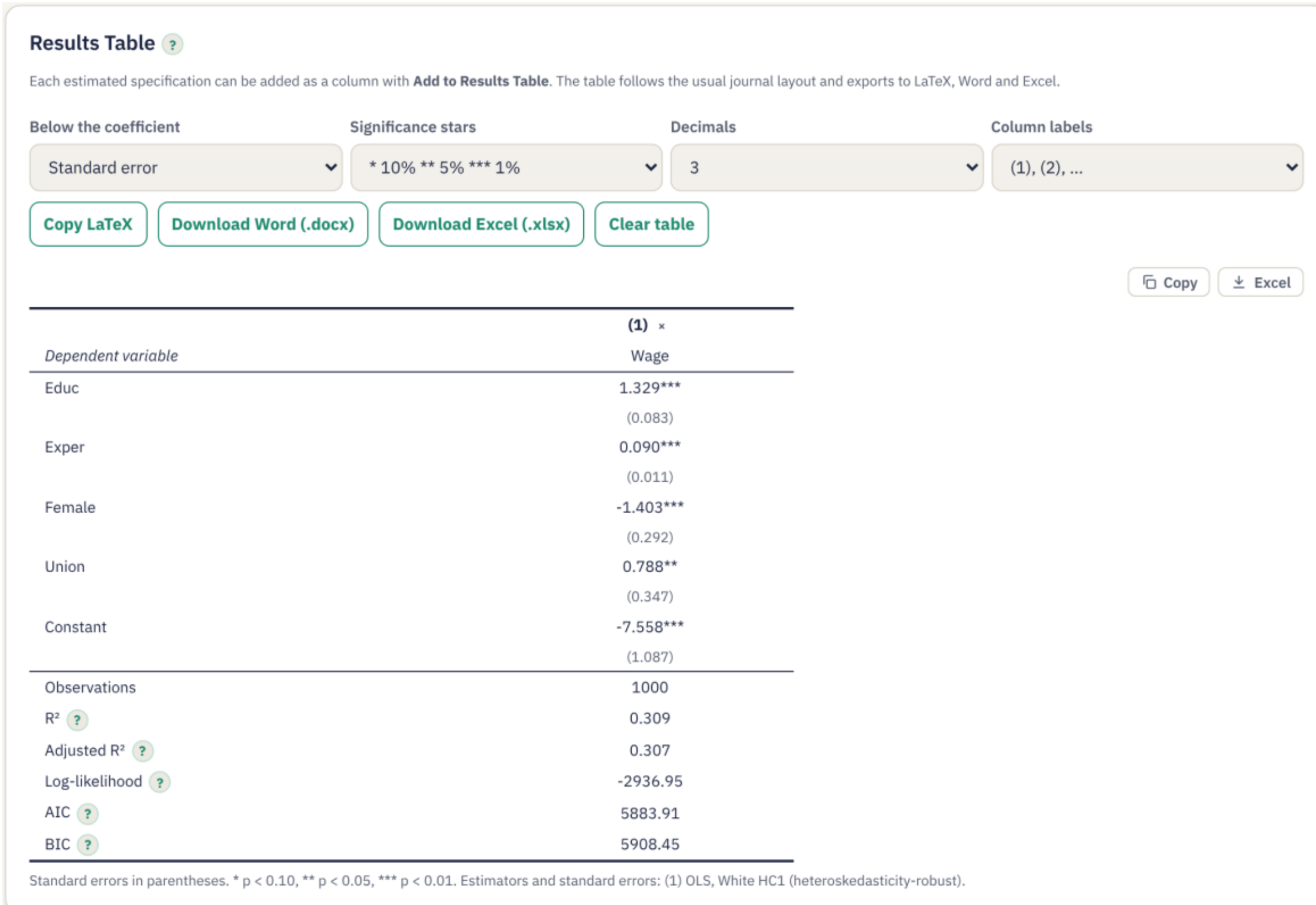

| | (1) |
|---|---|
| *Dependent variable* | Wage |
| Educ | 1.329*** |
| | (0.083) |
| Exper | 0.090*** |
| | (0.011) |
| Female | -1.403*** |
| | (0.292) |
| Union | 0.788** |
| | (0.347) |
| Constant | -7.558*** |
| | (1.087) |
| Observations | 1000 |
| $R^2$ | 0.309 |
| Adjusted $R^2$ | 0.307 |
| Log-likelihood | -2936.95 |
| AIC | 5883.91 |
| BIC | 5908.45 |

Standard errors in parentheses. * p < 0.10, ** p < 0.05, *** p < 0.01. Estimators and standard errors: (1) OLS, White HC1 (heteroskedasticity-robust).

**Figure 4.** The Results Table: estimated models side by side in journal layout, exported to LaTeX, Word or Excel.

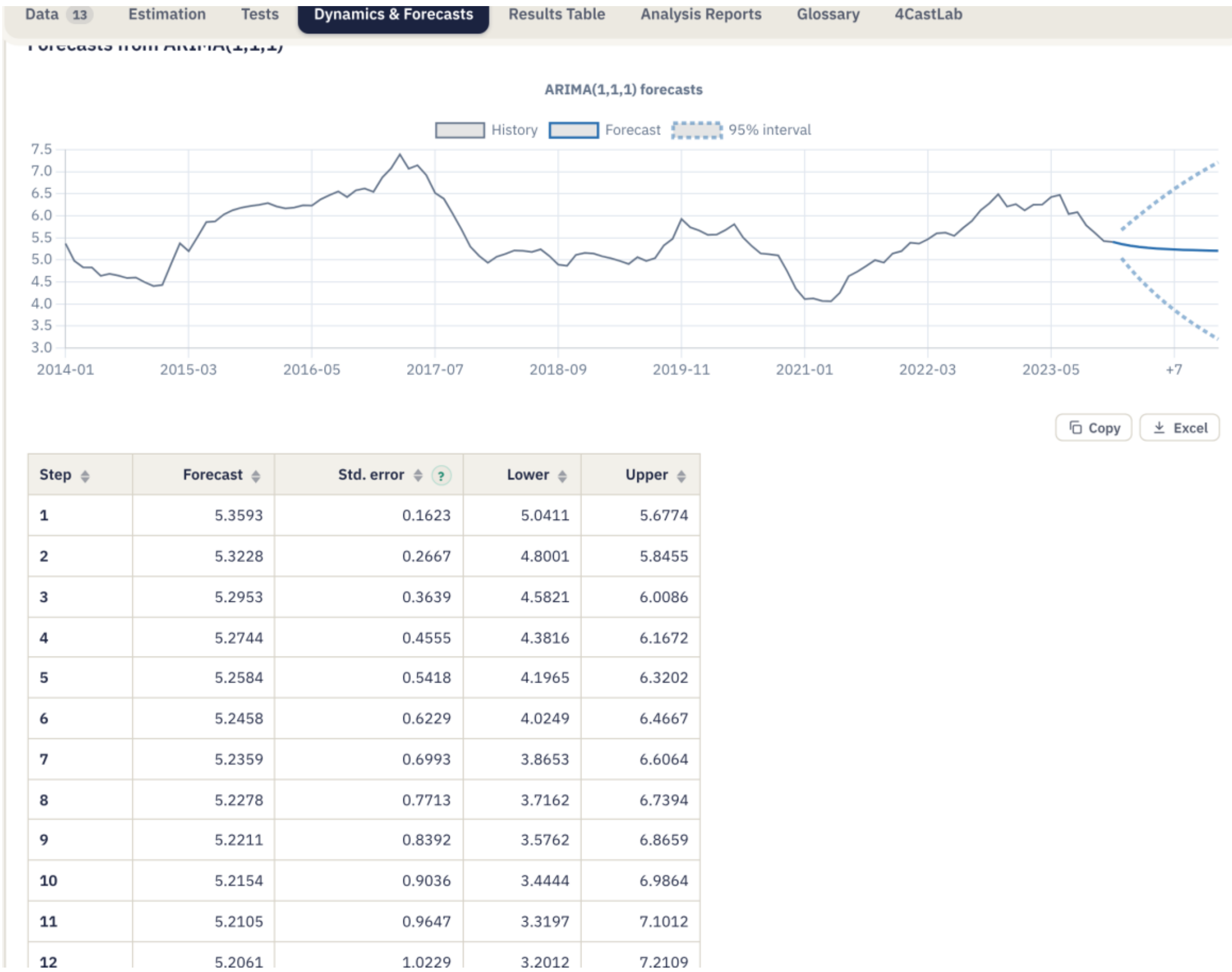

| Step | Forecast | Std. error | Lower | Upper |
|---|---|---|---|---|
| 1 | 5.3593 | 0.1623 | 5.0411 | 5.6774 |
| 2 | 5.3228 | 0.2667 | 4.8001 | 5.8455 |
| 3 | 5.2953 | 0.3639 | 4.5821 | 6.0086 |
| 4 | 5.2744 | 0.4555 | 4.3816 | 6.1672 |
| 5 | 5.2584 | 0.5418 | 4.1965 | 6.3202 |
| 6 | 5.2458 | 0.6229 | 4.0249 | 6.4667 |
| 7 | 5.2359 | 0.6993 | 3.8653 | 6.6064 |
| 8 | 5.2278 | 0.7713 | 3.7162 | 6.7394 |
| 9 | 5.2211 | 0.8392 | 3.5762 | 6.8659 |
| 10 | 5.2154 | 0.9036 | 3.4444 | 6.9864 |
| 11 | 5.2105 | 0.9647 | 3.3197 | 7.1012 |
| 12 | 5.2061 | 1.0229 | 3.2012 | 7.2109 |

**Figure 5.** Dynamics & Forecasts: the ARIMA(1,1,1) forecast of unemployment with its 95% interval.

# 6. Uses in research and teaching

For research and professional work, the Expert view, nested cross-validation, the pairwise tests, the reports and the Python export allow a model built in the app to be described and reproduced in a paper or a technical report. In teaching, the Core view, the plain-language summaries and the built-in examples support courses on machine learning, econometrics and data analytics, and three exercises illustrate the methodology: *benchmark first*, as in Section 5.2; *leakage by example*, where the same model is run with a random split and with an embargoed time split to show how look-ahead bias inflates accuracy; and *costs and thresholds*, where students set misclassification costs on the credit data and see the optimal threshold, the confusion matrix and the expected cost on the ROC panel. Two Quick Start videos and a written guide introduce the Forecasting and Estimation modes in a few minutes each. Doctoral students of the E|M|E|F Lab tested successive versions on their own datasets; their reports led to corrections in panel handling, benchmark choice and date parsing.

# 7. Limitations and future work

Because it runs in a browser engine, 4Cast Lab suits datasets of thousands to tens of thousands of rows rather than big data, and kernel methods are slower still because of the Gram matrix. Full tuning of all learners on 1,000 rows takes about two minutes. Random forest, boosting and the neural network agree with scikit-learn only approximately, since solvers and random number streams differ. Classification is currently binary, and the predictive-accuracy tests are applied to the Test Set only.

Planned work includes multi-class classification, rolling-origin evaluation of multi-step forecasts, and data-snooping-robust comparison of many models through White's (2000) Reality Check, Hansen's (2005) SPA test and the Model Confidence Set of Hansen, Lunde and Nason (2011).

## 8. Availability and citation

4Cast Lab is free software under the MIT license. The source code, releases, a Quick Start guide and the Quick Start videos are linked from https://github.com/PerryGogas/4Cast-Lab, and each release is archived on Zenodo. Users are asked to cite the software through its concept DOI, which always resolves to the latest version: 10.5281/zenodo.22818865.

*Suggested citation:* Gogas, P. (2026). 4Cast Lab: Open-source app for machine learning and statistical analyses (Version 4.6.2) [Computer software]. E|M|E|F Lab, Democritus University of Thrace. https://doi.org/10.5281/zenodo.22818865

Declaration of generative AI use. During the preparation of this work the author used Claude (Anthropic) to help draft and edit the text, to add screenshots of the app, to run the validation tests against reference software, and to identify and correct software defects. The author reviewed and edited all content, verified the results, and takes full responsibility for the publication.

# Appendix A. Screenshots (version 4.6.2)

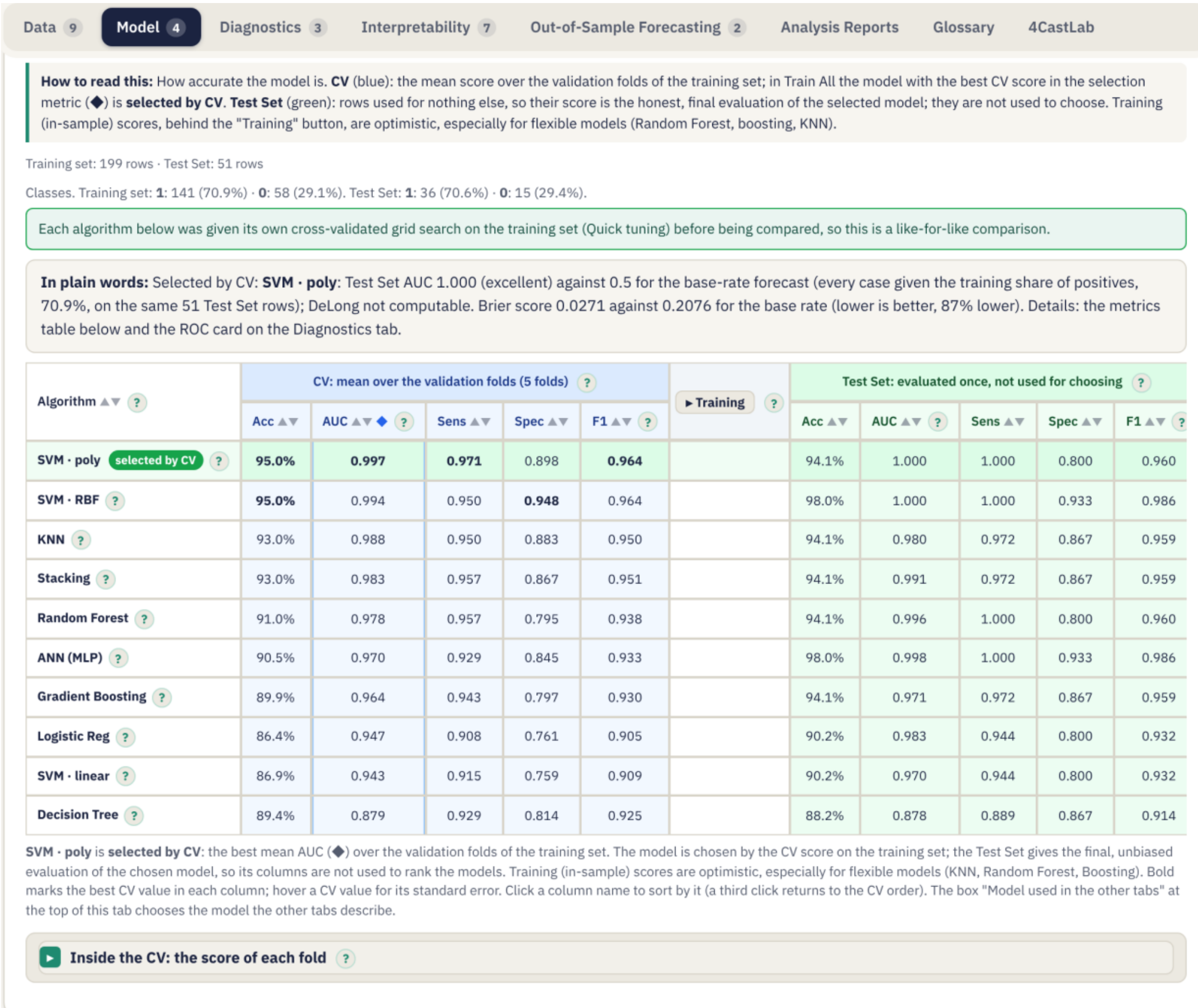
Data 9 | Model 4 | Diagnostics 3 | Interpretability 7 | Out-of-Sample Forecasting 2 | Analysis Reports | Glossary | 4CastLab

**How to read this:** How accurate the model is. **CV** (blue): the mean score over the validation folds of the training set; in Train All the model with the best CV score in the selection metric (◆) is **selected by CV**. **Test Set** (green): rows used for nothing else, so their score is the honest, final evaluation of the selected model; they are not used to choose. Training (in-sample) scores, behind the "Training" button, are optimistic, especially for flexible models (Random Forest, boosting, KNN).

Training set: 199 rows · Test Set: 51 rows

Classes. Training set: **1**: 141 (70.9%) · **0**: 58 (29.1%). Test Set: **1**: 36 (70.6%) · **0**: 15 (29.4%).

Each algorithm below was given its own cross-validated grid search on the training set (Quick tuning) before being compared, so this is a like-for-like comparison.

**In plain words:** Selected by CV: **SVM · poly**: Test Set AUC 1.000 (excellent) against 0.5 for the base-rate forecast (every case given the training share of positives, 70.9%, on the same 51 Test Set rows); DeLong not computable. Brier score 0.0271 against 0.2076 for the base rate (lower is better, 87% lower). Details: the metrics table below and the ROC card on the Diagnostics tab.

| Algorithm ▲▼ ? | CV: mean over the validation folds (5 folds) ? | | | | | ▸ Training ? | Test Set: evaluated once, not used for choosing ? | | | | |
|---|---|---|---|---|---|---|---|---|---|---|---|
| | Acc ▲▼ | AUC ▲▼ ◆ ? | Sens ▲▼ | Spec ▲▼ | F1 ▲▼ ? | | Acc ▲▼ | AUC ▲▼ ? | Sens ▲▼ | Spec ▲▼ | F1 ▲▼ ? |
| SVM · poly selected by CV ? | **95.0%** | **0.997** | **0.971** | 0.898 | **0.964** | | 94.1% | 1.000 | 1.000 | 0.800 | 0.960 |
| SVM · RBF ? | **95.0%** | 0.994 | 0.950 | **0.948** | 0.964 | | 98.0% | 1.000 | 1.000 | 0.933 | 0.986 |
| KNN ? | 93.0% | 0.988 | 0.950 | 0.883 | 0.950 | | 94.1% | 0.980 | 0.972 | 0.867 | 0.959 |
| Stacking ? | 93.0% | 0.983 | 0.957 | 0.867 | 0.951 | | 94.1% | 0.991 | 0.972 | 0.867 | 0.959 |
| Random Forest ? | 91.0% | 0.978 | 0.957 | 0.795 | 0.938 | | 94.1% | 0.996 | 1.000 | 0.800 | 0.960 |
| ANN (MLP) ? | 90.5% | 0.970 | 0.929 | 0.845 | 0.933 | | 98.0% | 0.998 | 1.000 | 0.933 | 0.986 |
| Gradient Boosting ? | 89.9% | 0.964 | 0.943 | 0.797 | 0.930 | | 94.1% | 0.971 | 0.972 | 0.867 | 0.959 |
| Logistic Reg ? | 86.4% | 0.947 | 0.908 | 0.761 | 0.905 | | 90.2% | 0.983 | 0.944 | 0.800 | 0.932 |
| SVM · linear ? | 86.9% | 0.943 | 0.915 | 0.759 | 0.909 | | 90.2% | 0.970 | 0.944 | 0.800 | 0.932 |
| Decision Tree ? | 89.4% | 0.879 | 0.929 | 0.814 | 0.925 | | 88.2% | 0.878 | 0.889 | 0.867 | 0.914 |

**SVM · poly** is **selected by CV**: the best mean AUC (◆) over the validation folds of the training set. The model is chosen by the CV score on the training set; the Test Set gives the final, unbiased evaluation of the chosen model, so its columns are not used to rank the models. Training (in-sample) scores are optimistic, especially for flexible models (KNN, Random Forest, Boosting). Bold marks the best CV value in each column; hover a CV value for its standard error. Click a column name to sort by it (a third click returns to the CV order). The box "Model used in the other tabs" at the top of this tab chooses the model the other tabs describe.

▸ Inside the CV: the score of each fold ?

**Figure A1.** Train All Models & Compare (Credit Default): every algorithm tuned the same way; the model with the best CV score is marked "selected by CV".

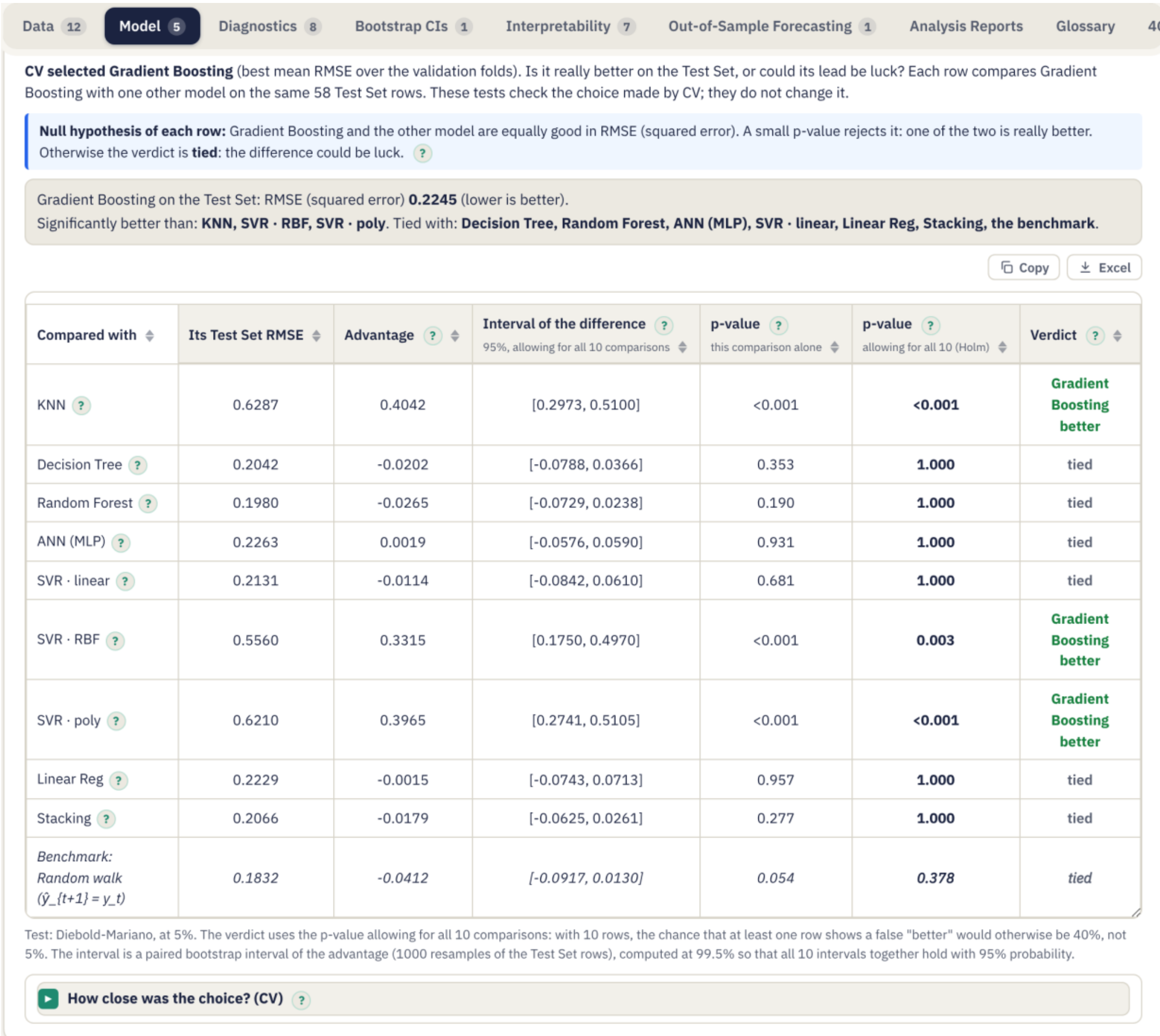

Data 12 | **Model 5** | Diagnostics 8 | Bootstrap CIs 1 | Interpretability 7 | Out-of-Sample Forecasting 1 | Analysis Reports | Glossary | 4C

**CV selected Gradient Boosting** (best mean RMSE over the validation folds). Is it really better on the Test Set, or could its lead be luck? Each row compares Gradient Boosting with one other model on the same 58 Test Set rows. These tests check the choice made by CV; they do not change it.

**Null hypothesis of each row:** Gradient Boosting and the other model are equally good in RMSE (squared error). A small p-value rejects it: one of the two is really better. Otherwise the verdict is **tied**: the difference could be luck. ?

Gradient Boosting on the Test Set: RMSE (squared error) **0.2245** (lower is better).
Significantly better than: **KNN, SVR · RBF, SVR · poly**. Tied with: **Decision Tree, Random Forest, ANN (MLP), SVR · linear, Linear Reg, Stacking, the benchmark**.

Copy | Excel

| Compared with | Its Test Set RMSE | Advantage ? | Interval of the difference ? 95%, allowing for all 10 comparisons | p-value ? this comparison alone | p-value ? allowing for all 10 (Holm) | Verdict ? |
|---|---|---|---|---|---|---|
| KNN ? | 0.6287 | 0.4042 | [0.2973, 0.5100] | <0.001 | **<0.001** | **Gradient Boosting better** |
| Decision Tree ? | 0.2042 | -0.0202 | [-0.0788, 0.0366] | 0.353 | **1.000** | tied |
| Random Forest ? | 0.1980 | -0.0265 | [-0.0729, 0.0238] | 0.190 | **1.000** | tied |
| ANN (MLP) ? | 0.2263 | 0.0019 | [-0.0576, 0.0590] | 0.931 | **1.000** | tied |
| SVR · linear ? | 0.2131 | -0.0114 | [-0.0842, 0.0610] | 0.681 | **1.000** | tied |
| SVR · RBF ? | 0.5560 | 0.3315 | [0.1750, 0.4970] | <0.001 | **0.003** | **Gradient Boosting better** |
| SVR · poly ? | 0.6210 | 0.3965 | [0.2741, 0.5105] | <0.001 | **<0.001** | **Gradient Boosting better** |
| Linear Reg ? | 0.2229 | -0.0015 | [-0.0743, 0.0713] | 0.957 | **1.000** | tied |
| Stacking ? | 0.2066 | -0.0179 | [-0.0625, 0.0261] | 0.277 | **1.000** | tied |
| *Benchmark: Random walk (ŷ_{t+1} = y_t)* | *0.1832* | *-0.0412* | *[-0.0917, 0.0130]* | *0.054* | ***0.378*** | *tied* |

Test: Diebold-Mariano, at 5%. The verdict uses the p-value allowing for all 10 comparisons: with 10 rows, the chance that at least one row shows a false "better" would otherwise be 40%, not 5%. The interval is a paired bootstrap interval of the advantage (1000 resamples of the Test Set rows), computed at 99.5% so that all 10 intervals together hold with 95% probability.

▶ **How close was the choice? (CV)** ?

**Figure A2.** Forecasting example: the model selected by CV against every alternative and the random walk (Diebold-Mariano, Holm).

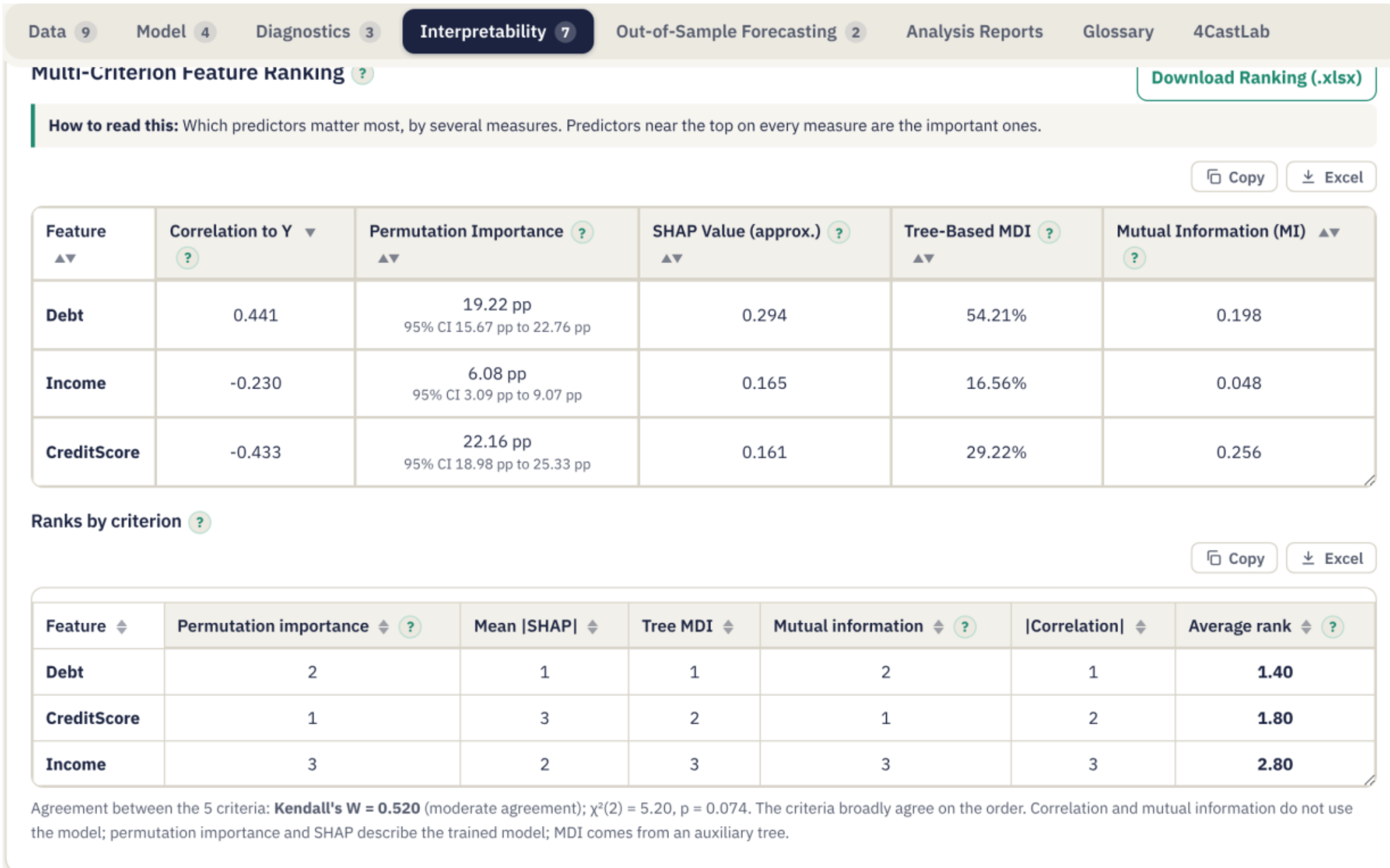

Data 9 | Model 4 | Diagnostics 3 | Interpretability 7 | Out-of-Sample Forecasting 2 | Analysis Reports | Glossary | 4CastLab

Multi-Criterion Feature Ranking

Download Ranking (.xlsx)

**How to read this:** Which predictors matter most, by several measures. Predictors near the top on every measure are the important ones.

| Feature | Correlation to Y | Permutation Importance | SHAP Value (approx.) | Tree-Based MDI | Mutual Information (MI) |
|---|---|---|---|---|---|
| Debt | 0.441 | 19.22 pp<br>95% CI 15.67 pp to 22.76 pp | 0.294 | 54.21% | 0.198 |
| Income | -0.230 | 6.08 pp<br>95% CI 3.09 pp to 9.07 pp | 0.165 | 16.56% | 0.048 |
| CreditScore | -0.433 | 22.16 pp<br>95% CI 18.98 pp to 25.33 pp | 0.161 | 29.22% | 0.256 |

Ranks by criterion

| Feature | Permutation importance | Mean \|SHAP\| | Tree MDI | Mutual information | \|Correlation\| | Average rank |
|---|---|---|---|---|---|---|
| Debt | 2 | 1 | 1 | 2 | 1 | **1.40** |
| CreditScore | 1 | 3 | 2 | 1 | 2 | **1.80** |
| Income | 3 | 2 | 3 | 3 | 3 | **2.80** |

Agreement between the 5 criteria: **Kendall's W = 0.520** (moderate agreement); $\chi^2(2) = 5.20$, p = 0.074. The criteria broadly agree on the order. Correlation and mutual information do not use the model; permutation importance and SHAP describe the trained model; MDI comes from an auxiliary tree.

**Figure A3.** Interpretability: predictors ranked by five criteria, with Kendall's W for their agreement.

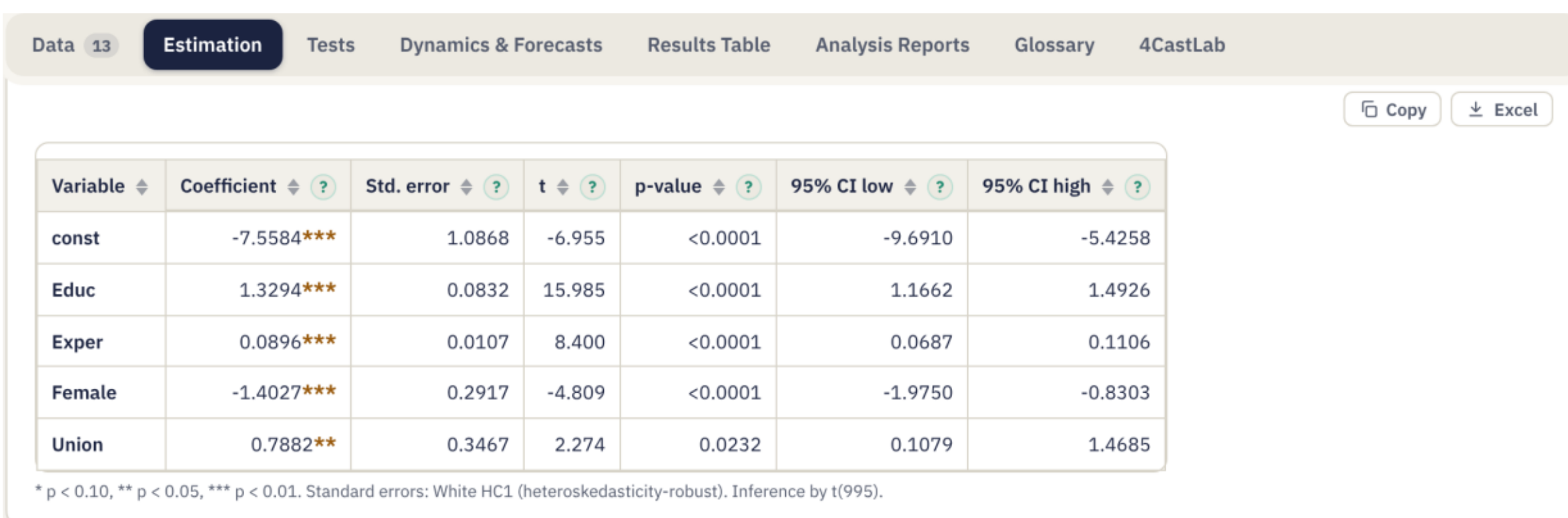

Data 13 | Estimation | Tests | Dynamics & Forecasts | Results Table | Analysis Reports | Glossary | 4CastLab

| Variable | Coefficient | Std. error | t | p-value | 95% CI low | 95% CI high |
|---|---|---|---|---|---|---|
| const | -7.5584*** | 1.0868 | -6.955 | <0.0001 | -9.6910 | -5.4258 |
| Educ | 1.3294*** | 0.0832 | 15.985 | <0.0001 | 1.1662 | 1.4926 |
| Exper | 0.0896*** | 0.0107 | 8.400 | <0.0001 | 0.0687 | 0.1106 |
| Female | -1.4027*** | 0.2917 | -4.809 | <0.0001 | -1.9750 | -0.8303 |
| Union | 0.7882** | 0.3467 | 2.274 | 0.0232 | 0.1079 | 1.4685 |

* p < 0.10, ** p < 0.05, *** p < 0.01. Standard errors: White HC1 (heteroskedasticity-robust). Inference by t(995).

**Figure A4.** Estimation mode: OLS coefficients with White HC1 standard errors, t statistics, p-values and 95% intervals.

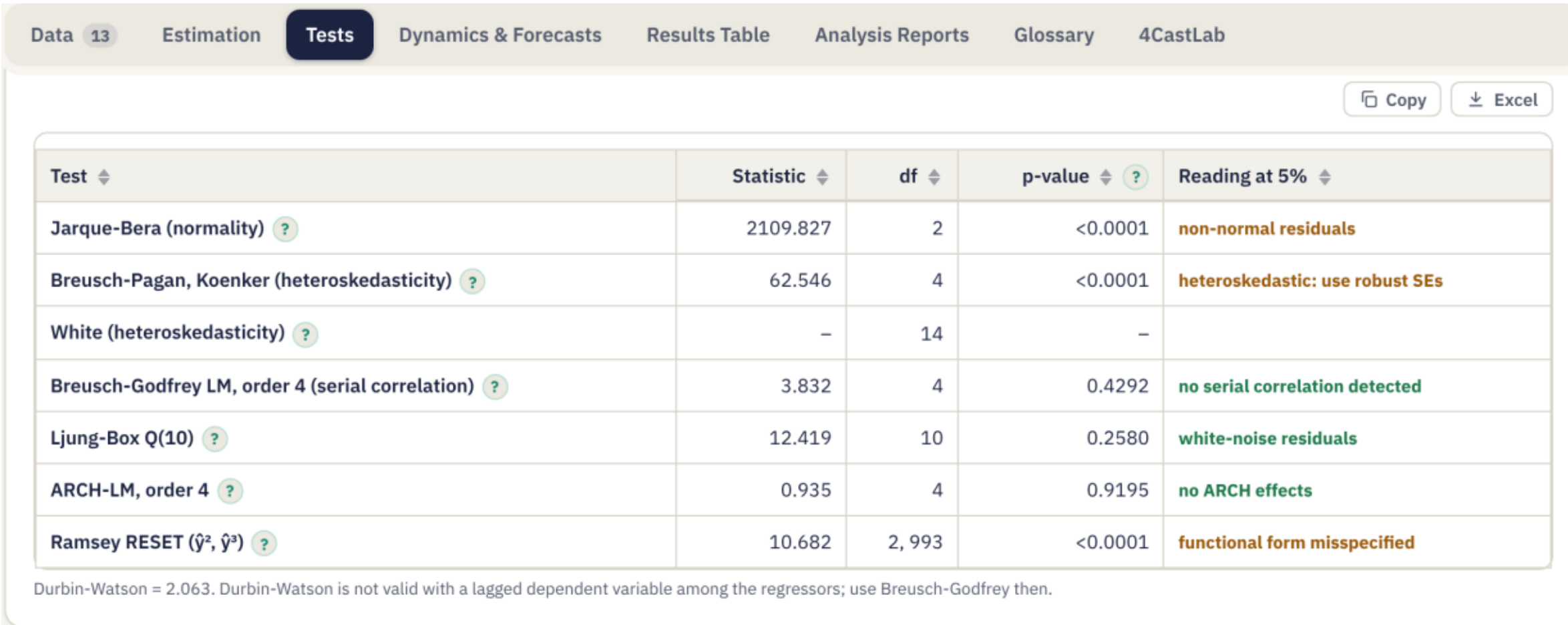

| Test | Statistic | df | p-value | Reading at 5% |
|---|---|---|---|---|
| Jarque-Bera (normality) | 2109.827 | 2 | <0.0001 | non-normal residuals |
| Breusch-Pagan, Koenker (heteroskedasticity) | 62.546 | 4 | <0.0001 | heteroskedastic: use robust SEs |
| White (heteroskedasticity) | – | 14 | – | |
| Breusch-Godfrey LM, order 4 (serial correlation) | 3.832 | 4 | 0.4292 | no serial correlation detected |
| Ljung-Box Q(10) | 12.419 | 10 | 0.2580 | white-noise residuals |
| ARCH-LM, order 4 | 0.935 | 4 | 0.9195 | no ARCH effects |
| Ramsey RESET ($\hat{y}^2$, $\hat{y}^3$) | 10.682 | 2, 993 | <0.0001 | functional form misspecified |

Durbin-Watson = 2.063. Durbin-Watson is not valid with a lagged dependent variable among the regressors; use Breusch-Godfrey then.

**Figure A5.** The Tests tab: residual diagnostics and specification tests, each with a plain reading at 5%.

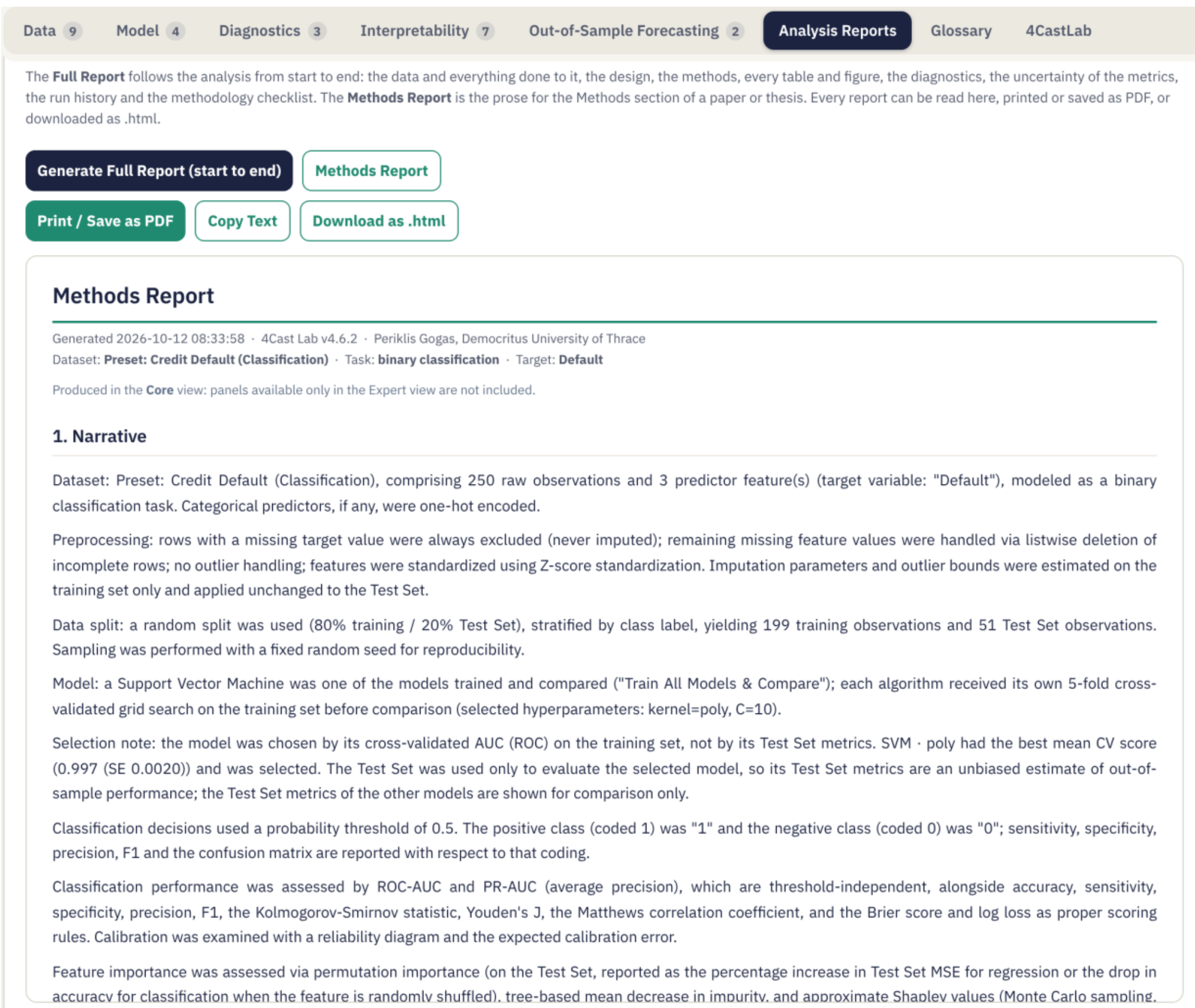

## Methods Report

Generated 2026-10-12 08:33:58 · 4Cast Lab v4.6.2 · Periklis Gogas, Democritus University of Thrace

Dataset: **Preset: Credit Default (Classification)** · Task: **binary classification** · Target: **Default**

Produced in the **Core** view: panels available only in the Expert view are not included.

### 1. Narrative

Dataset: Preset: Credit Default (Classification), comprising 250 raw observations and 3 predictor feature(s) (target variable: "Default"), modeled as a binary classification task. Categorical predictors, if any, were one-hot encoded.

Preprocessing: rows with a missing target value were always excluded (never imputed); remaining missing feature values were handled via listwise deletion of incomplete rows; no outlier handling; features were standardized using Z-score standardization. Imputation parameters and outlier bounds were estimated on the training set only and applied unchanged to the Test Set.

Data split: a random split was used (80% training / 20% Test Set), stratified by class label, yielding 199 training observations and 51 Test Set observations. Sampling was performed with a fixed random seed for reproducibility.

Model: a Support Vector Machine was one of the models trained and compared ("Train All Models & Compare"); each algorithm received its own 5-fold cross-validated grid search on the training set before comparison (selected hyperparameters: kernel=poly, C=10).

Selection note: the model was chosen by its cross-validated AUC (ROC) on the training set, not by its Test Set metrics. SVM · poly had the best mean CV score (0.997 (SE 0.0020)) and was selected. The Test Set was used only to evaluate the selected model, so its Test Set metrics are an unbiased estimate of out-of-sample performance; the Test Set metrics of the other models are shown for comparison only.

Classification decisions used a probability threshold of 0.5. The positive class (coded 1) was "1" and the negative class (coded 0) was "0"; sensitivity, specificity, precision, F1 and the confusion matrix are reported with respect to that coding.

Classification performance was assessed by ROC-AUC and PR-AUC (average precision), which are threshold-independent, alongside accuracy, sensitivity, specificity, precision, F1, the Kolmogorov-Smirnov statistic, Youden's J, the Matthews correlation coefficient, and the Brier score and log loss as proper scoring rules. Calibration was examined with a reliability diagram and the expected calibration error.

Feature importance was assessed via permutation importance (on the Test Set, reported as the percentage increase in Test Set MSE for regression or the drop in accuracy for classification when the feature is randomly shuffled), tree-based mean decrease in impurity, and approximate Shapley values (Monte Carlo sampling,

**Figure A6.** The Methods Report: prose for the Methods section of a paper or thesis, generated for each run.